\documentclass[noshowpacs,amsmath,twocolumn,aps,prb]{revtex4-1}

\usepackage{setspace}
\usepackage{amsmath}
\usepackage{graphicx}
\usepackage[nearskip,margin = 0pt]{subfig}
\usepackage{subfig}
\usepackage{verbatim}
\usepackage{amsfonts}
\usepackage{amssymb}
\usepackage{epstopdf} 
\usepackage{ulem}
\usepackage{xcolor}
\DeclareGraphicsExtensions{.pdf,.eps,.png,.jpg,.mps}
\usepackage{ragged2e}
\usepackage{hyperref}
\hypersetup{
    colorlinks=true,
    linkcolor=blue,
    citecolor=blue,
    filecolor=blue,      
    urlcolor=blue,
}

\begin{document}

\title{Kerr nonlinearity, self-injection locking and correlation in a microresonator}

\author{ Andrey Matsko$^{*,1}$, Abdelkrim El Amili$^2$, and Lute Maleki$^{2}$,  \\
$^1$Jet Propulsion Laboratory, California Institute of Technology, 4800 Oak Grove Drive, Pasadena, California, 91109-8099, USA. \\
$^2$OEwaves Inc., 465 North Halstead Street, Suite 140, Pasadena, CA, 91107, USA; \\
$^*$Corresponding author: andrey.b.matsko@jpl.nasa.gov}

\maketitle

{\bf Production of entangled photon pairs is important in secure communication systems, quantum computing, and fundamental physics experiments. Achieving efficient generation of such photon pairs with low-loss parametric oscillators is a key objective in advancing integrated quantum technologies. However, spatially separating the generated photons while preserving their entanglement represents a significant technical challenge. In this work, we demonstrate nonlinear generation of correlated optical harmonics based on non-degenerate four-wave mixing with an optimally pumped optical microcavity with Kerr nonlinearity. The phase matching of the process is achieved with self-injection locked lasers producing parametric oscillation while locked to two different modes of the microresonator. This condition is reminiscent of slow-light technique developed for coherent atomic systems. The experimental design, utilizing counterpropagating light from two self-injection locked lasers, also effectively addresses the challenge of spatial separation of the generated harmonics. We validate the theoretical predictions using two self-injection locked semiconductor lasers integrated with a crystalline whispering gallery mode resonator with optimized spectral structure.  \copyright 2025 All Rights Reserved
}\\

\noindent{\textbf{Introduction.}} In the mid-1990's research on optical whispering gallery mode resonators (WGMRs), which had begun decades earlier \cite{matsko06jstqe,chiasera10lpr}, culminated in new applications based on new resonator morphologies and materials.  A major milestone corresponding to the quality factor (Q-) exceeding 10 billion \cite{gorodetsky96ol,savchenkov07oe}, was also reached to pave the way for new applications made possible by the ability to excite optical nonlinearities with modest input powers. Interest in WGMRs surged in the early 2000s, largely propelled by the discovery of optical frequency combs \cite{delhaye07n,savchenkov08prl,kippenberg11s}. These so-called microcombs were generated through cascaded four-wave mixing (FWM), a nonlinear optical process enabled through interaction between the Kerr nonlinearity of the resonator material and laser light pumping a resonator mode, and were demonstrated with both bulk and on-chip ring WGMRs.  

By 2010, a new paradigm in laser-resonator interactions emerged with the realization of self-injection locking (SIL) of semiconductor lasers to a WGM \cite{liang10ol}. In this process, which was introduced much earlier with electronic oscillators \cite{ohta68ecj} as well as semiconductor lasers integrated with Fabry-Perot cavities \cite{dahmani87ol}, the interaction between the laser and the resonator remains within the linear regime. Resonant Rayleigh scattering from the high-quality-factor optical mode feeds a portion of the reflected light back into the laser cavity \cite{gorodetsky00josab}. This feedback mechanism significantly reduces the phase and amplitude noise of the laser, leading to improved spectral purity and frequency stability \cite{liang15nc}. SIL was also found to significantly simplify  generation of combs \cite{voloshin21nc} and as a result, commercially packaged Kerr microcomb oscillators were created \cite{maleki11np}. The combination of SIL and microcombs has gained widespread adoption in integrated photonics, where both on-chip and off-chip lasers are coupled to on-chip resonators \cite{kondratiev23fp}. This coupling dramatically suppresses frequency and amplitude noise of the semiconductor laser and enhances the stability of frequency combs  \cite{liang15nc1}.

The combination of SIL and microcombs \cite{liang15nc,shen20n} has enabled transformative applications across diverse fields, including high-speed optical communications \cite{geng22ol}, ultra-precise spectroscopy \cite{pisque19np,savchenkov20sr}, precision metrology \cite{voloshin24arxiv}, and emerging quantum technologies \cite{heim25arxiv}. As photonic integration continues to evolve, these innovations are expected to play an increasingly critical role in next-generation optical systems.  

More recently, researchers have begun investigating the effects of light from two lasers on WGMRs to understand how the presence of dual light fields alters the dynamics of nonlinear interactions within the resonators \cite{hansson14pra, wang16sr, suzuki18pj, taheri22nc, wildi23apl, moille24np}.  A particularly striking recent development in this field is the realization of a time crystal within a WGMR \cite{taheri22nc}, emerging from the interplay between two light fields exciting distinct resonator modes. This discovery highlights the profound impact of multi-laser interactions with WGMRs, opening new avenues for research in nonlinear optics and quantum photonics.

Multiple experiments have demonstrating the use of two lasers coupled to a microresonator. SIL of two lasers to different modes of a WGMR was initially studied in \cite{liang15nc1}. Injection locking was shown to reduce the relative noise of the lasers by subtracting the correlated noise components, resulting in the same close-in phase noise in both lasers related to the frequency jitter of the resonator modes. The RF beat note produced by the lasers on a fast photodiode also had reduced noise because the common noise contribution resulting from the jitter is removed at the photodiode. Simultaneous SIL of two vertical cavity surface emitting lasers (VCSELs) with copropagating emission to two different resonances of a single WGMR made of Hydex glass was experimentally demonstrated more recently \cite{jiang21oe}.  Subsequently, dual-laser SIL of two copropagating multifrequency laser diodes was demonstrated in different modes of an integrated silicon nitride microresonator \cite{chermoshentsev22oe}. Simultaneous spectrum collapse of both lasers, as well as linewidth narrowing and high-frequency noise suppression, together with strong nonlinear interaction of the two fields with each other, was also observed. 

In this paper, we explore the interaction between \textit{two lasers and a WGMR}, emphasizing the three-way interplay between the resonator and both lasers, rather than simply the interaction of \textit{two laser-generated light fields} with the resonator. This distinction is critical, as previous studies have already shown that light from two lasers -- whether at identical or different wavelengths -- can result in nonlinear interactions with a WGMR, affecting, for example, both the formation and stability of optical frequency combs \cite{hansson14pra, wang16sr, suzuki18pj, taheri22nc, wildi23apl, moille24np}. 

Since FWM and cascaded FWM are essentially nonlinear processes, they are used to generate non-classical light \cite{chembo16pra}. Bichromatically pumped microresonators have also been used for the generation of squeezed light,  but only through \textit{degenerate} FWM process observed in a copropagating light geometry \cite{zhao20prl}. Multiple quantum correlated harmonic generation is frequently observed in this configuration as multiple modes satisfy the phase matching conditions of the process \cite{okawachi15ol,hu17pr,seifoory22pra}. However, the presence of multiple harmonics is not always practically advantageous. 

It is desirable to obtain a canonical FWM process in which only two harmonics are generated when either one or two pumps are applied. The possibility of such a process was proven for a photonic crystal resonator pumped with a single laser \cite{marty21np}. Similarly, a nearly ideal FWM process was observed in a monochromatically pumped microcavity in the co-propagating regime \cite{dutt15pra}. A configuration suitable for the generation of significantly nondegenerate harmonics with a single pump was also demonstrated \cite{matsko16ol}. Although, as expected, the generated harmonics were quantum correlated, their separation required usage of optical filters that can introduce undesirable insertion loss; a counterpropagating geometry would be more effective here. 

In the case of two counterpropagating light fields resonant with cavity modes, the FWM process is usually not phase-matched; so two counter-propagating pumps cannot generate harmonics confined in the same resonator. For instance, a counter-propagating pump can be used for thermal control of the cavity but without a direct influence on  Kerr comb generation in the opposite direction \cite{shou19lsa}. Counterpropagating pumps also result in symmetry-breaking not associated with the FWM process \cite{delbino17sr}. 

In the degenerate or nearly-degenerate FWM case, interaction between counterpropagating cavity modes becomes possible. Energy oscillations between counterpropagating cavity modes are supported in the nearly degenerate case via resonant Rayleigh scattering \cite{yoshiki15oe,matsko17ol}.  Recently, it was shown that bichromatically pumped FWM is phase matched when the generated signals are counterpropagating and coinciding in frequency with the pumps \cite{RodríguezBecerra24ao}. As the first result of this study, we show how to achieve phase-matched FWM using two nondegenerate counterpropagating pumps. In our case, the frequency separation between the generated harmonics  is not related to the frequency separation between the pumps. We demonstrate this effect experimentally using two SIL lasers.

In some coherent atomic media the phase matching takes advantage of the phenomenon of slow light that enables resonant FWM \cite{zibrov99prl,liu19prl}. It was shown a while ago that two counterpropagating pumps in a double-$\Lambda$ atomic level system result in efficient generation of quantum correlated photons. Similarly, in this work we demonstrate both theoretically and experimentally that a similar physical process can be established in a nonlinear microcavity with properly selected optical spectrum where group velocity of light can be freely manipulated.

The study presented here specifically investigates FWM when two lasers are self-injection locked to two separate modes of the same WGMR. This phenomenon involves a three-way interaction between the lasers and the resonator. Notably, this interaction is not limited to the excitation of Kerr nonlinearity by the light fields; it also includes the back-action of this nonlinear process on the dynamics of the lasers themselves. This back-action, experimentally observed for the first time, manifests as the modulation of light emitted by one SIL laser in response to the current modulation of the second SIL laser.  

Our experimental setup leverages a counterpropagating configuration for the interacting light fields, which also naturally facilitates spatial separation of the generated harmonics. The resulted FWM produces bright signal and idler sidebands, which may serve as robust sources of entangled photon pairs. The inherent spatial separation of generated fields is particularly advantageous for practical quantum optics applications as it simplifies photon extraction and routing within integrated photonic circuits. Therefore, by enabling efficient and high-brightness entangled photon generation with built-in spatial separation, our approach holds major promise for advancing quantum communication, quantum computing, and other quantum applications of technology.

\vspace{3 mm}

\noindent {\bf Results}

\noindent {\bf Phase matching conditions}

In this section we address the question of phase matching conditions related to the FWM process in a WGMR being pumped with two laser in both copropagating  and counterpropagating configurations. Generation of FWM requires energy and photon number conservation, namely:
\begin{equation} \label{energyconserve}
    \omega_{p1}+\omega_{p2} = \omega_{s1}+\omega_{s2},
\end{equation}
where $\omega_{p1}$ and $\omega_{p2}$ as well as $\omega_{s1}$ and $\omega_{s2}$ are frequencies of the pumps and the signals involved in the four-wave mixing process. Physically, this requirement means that the energy of the two  pump photons is attenuated by transfer of the energy to the emitted sideband photons. As a result, the signal and pump doublets must be symmetric with respect to some central frequency,
\begin{equation}
    \omega_{0} \equiv \frac{\omega_{p1}+\omega_{p2}}{2}=\frac{\omega_{s1}+\omega_{s2}}{2}.
\end{equation}

As a coherent nonlinear process, FWM is supported when both  the pump and the signal waves are phase-matched. The phase-matching condition can be expressed in terms of the four-point overlap integral. Specifically, if the field of each interacting mode can be represented as the product of time-dependent and spatial-dependent components 
\begin{equation}
    E_j=\Psi_j({\bf r}) c_j(t),
\end{equation}
the phase matching requires 
\begin{equation}
    \int_V \Psi_i({\bf r}) \Psi_j({\bf r}) \Psi_k({\bf r}) \Psi_l({\bf r}) d{\bf r} \ne 0.
\end{equation}

The phase matching conditions are different in the co- and counter-propagation cases of the waves. In the co-propagation 1D configuration one needs to achieve the condition:
\begin{equation} \label{comomentumconserve}
     k_{p1}+ k_{p2} =  k_{s1}+  k_{s2} 
\end{equation}
for the absolute values of the wave lectors. Conditions (\ref{energyconserve}) and (\ref{comomentumconserve}) are nearly complementary and their implementation depends on the group velocity dispersion (GVD) term of the system.  In the case of zero GVD the condition (\ref{comomentumconserve}) is always implemented since  $k_j = \omega_j n(\omega_j)/c$. The optimal performance of the system can be achieved when GVD is not negligible, but comparable with the nonlinear frequency shifts. 

In the case of the 1D counter-propagating pumps, the phase matching condition is
\begin{equation} \label{countermomentumconserve}
     k_{p2}-  k_{p1} =  k_{s2}-  k_{s1}. 
\end{equation}
This condition cannot be implemented for the strongly frequency nondegenerate case. For example, in the dispersion-less case (refractive index $n(\omega)=const$), it calls for 
\begin{equation} \label{countermomentumconserve1}
    \omega_{p1}- \omega_{p2} = \omega_{s2}- \omega_{s1}.
\end{equation}
This is possible, for instance, if $\omega_{p1}=\omega_{s2}$ and $\omega_{p2}=\omega_{s1}$, as was shown in the literature. This case is not very practical in a WGMR since the nonlinear frequency interaction is masked by the linear Rayleigh scattering when a pump and a generated signal propagate in opposite directions. 

\begin{figure}[htbp]
\centering
\includegraphics[width=8.5cm]{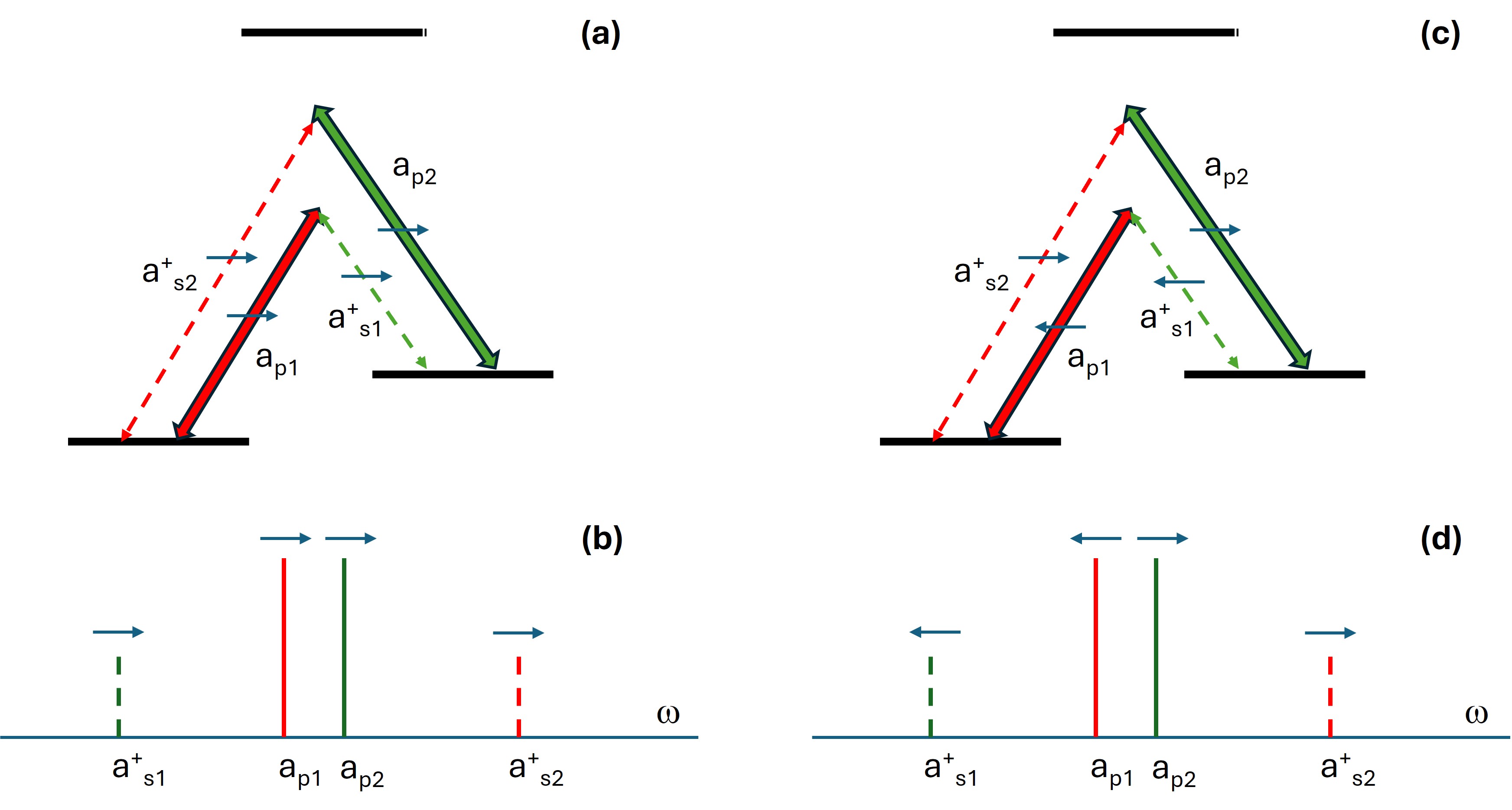}
\captionsetup{singlelinecheck=off, justification = Justified}
\caption{ {\bf Possible geometry of the FWM processes.}
Analogy between resonant atomic systems and resonant cavity-based photon pair generation schemes. 
(a) A double-lambda atomic configuration involving two co-propagating optical pump fields with orthogonal polarizations. This setup facilitates nonlinear interactions through atomic coherence.
(b) A resonant ring cavity-enhanced four-wave mixing process, where two co-propagating optical pump fields interact within the cavity. These pumps belong to distinct mode families of the ring cavity, enabling efficient parametric conversion and photon pair generation.
(c) A double-lambda atomic configuration with two counter-propagating optical pump fields that have orthogonal polarizations. This configuration modifies the phase-matching conditions and interaction dynamics compared to the co-propagating case.
(d) A resonant ring cavity-enhanced four-wave mixing process with two counter-propagating optical pumps. Similar to (b), the pumps belong to different mode families of the ring cavity, but the counter-propagating nature introduces additional phase-matching constraints and interaction symmetries. }
\label{figures:fig1} 
\end{figure}

To find a more general configuration for the phase matched modes it is useful to look at the analogies between the resonant FWM in an optical cavity and in an atomic vapor. As shown in Figs.~(\ref{figures:fig1}a) and (\ref{figures:fig1}c) two general FWM configurations can be supported by the atomic systems. In both  systems the pumps are orthogonally polarized and  co- \cite{mccormick08pra,pooser09oe} and counter-propagating \cite{zibrov99prl,liu19prl} processes can both be supported. 

\begin{figure}[htbp]
  \centering
\includegraphics[width=8.5cm]{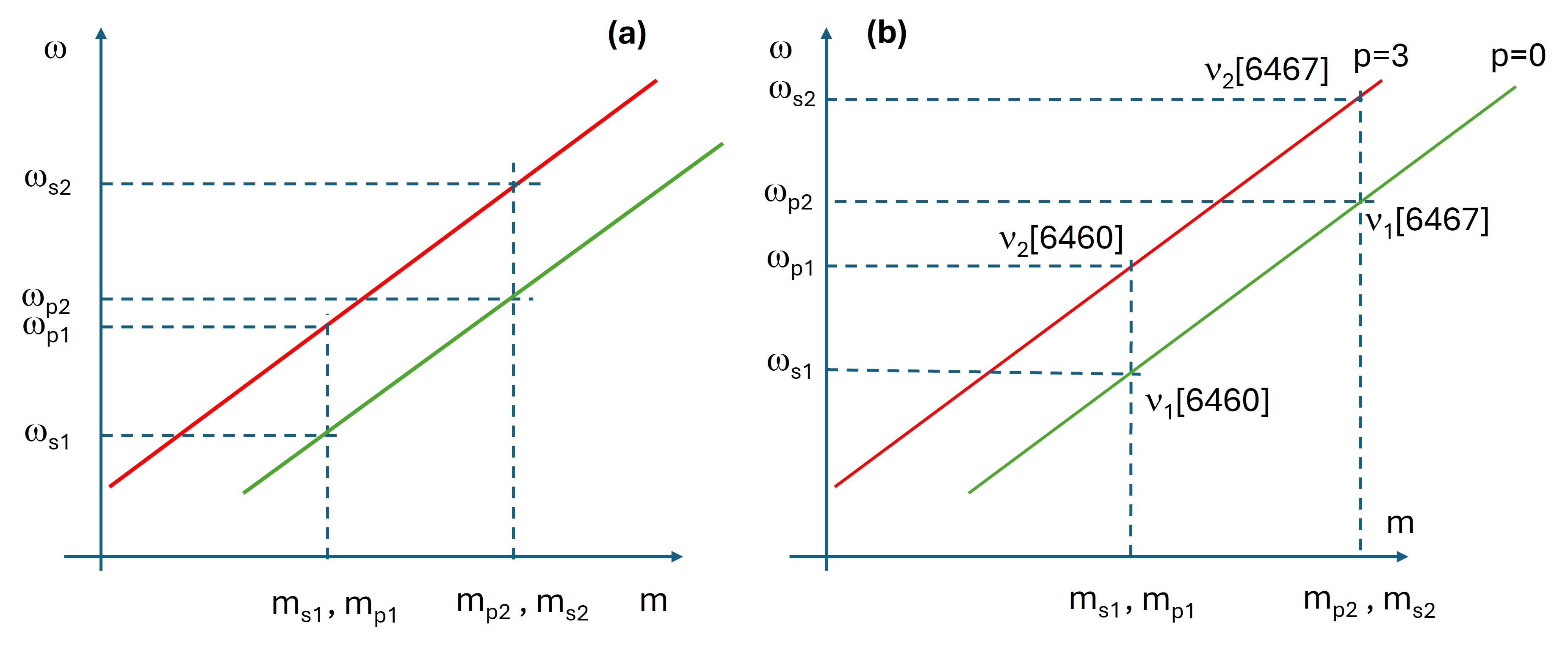}
\captionsetup{singlelinecheck=off, justification = Justified}
\caption{{\bf Dispersion relation supporting FWM.} (a) An illustration of the dispersion relations for two pairs of counter-propagating optical modes that satisfy phase-matching conditions supporting the process depicted in Figure~(\ref{figures:fig1}d). Here, $\omega_i$ represents the eigenfrequency of a mode, while 
$m_i$ denotes its azimuthal mode number. The color of the dispersion curves emphasizes that the modes belong to different mode families.
(b) A practical implementation of the mode configuration described in panel (a). The modes with eigenfrequencies $\nu_1$ and $\nu_2$ exist within the same optical microcavity made out of magnesium fluoride, whose spectrum is characterized by the quantum numbers 
(m,p,q). 
\label{figures:fig2} }
\end{figure}

To follow these analogies for our case, we assumed that the pumps and signals belong to two different mode families of a nonlinear resonator and considered the 3D case. A configuration of the modes is illustrated by Fig.~(\ref{figures:fig2}a). We demand the azimuthal mode numbers of the signal and pumps to coincide pairwise. 

For instance, it is possible to present the modes of the nonlinear cavity using a cylindrical coordinate model with axis $z$ coinciding with the resonator axis,
\begin{equation}
   \Psi \sim e^{\pm im\phi} e^{\pm i \int \beta(z) dz} R(\rho(z)),
\end{equation}
where $m$ is the azimuthal index corresponding to the number of field periods along the rim (ring) of the  resonator in the equatorial plane \cite{sumetsky04ol,demchenko13josab}. Using this model, we find that selecting two mode families with different quantum numbers $p$ and the same numbers $q$ it is possible to choose quantum numbers $m$ so that the mode structure shown in Fig.~(\ref{figures:fig1}a) is realized. A practical case for the spectrum of a MgF$_2$ WGM resonator characterized with $1.2$~mm horizontal curvature radius and $0.4$~mm vertical curvature radius is illustrated in Fig.~(\ref{figures:fig1}b). These modes have nonzero overlap integral and can support the FWM process. Importantly, there are no other signal pairs that can compete with the identified signal pair.  

Interestingly, the phase matching of the process can be directly linked to the slow and fast light phenomena in atomic vapor systems. The beat note speed is given by the ratio of the frequency difference and the wave number difference of the harmonics. Harmonics with the same frequency and different wave numbers create a stationary interference pattern \cite{savchenkov07pra}. The slope of the dispersion curve for a single mode family Fig.~(\ref{figures:fig2}) illustrates the group velocity of the mode family. The beat note of two pump lasers propagates with a speed smaller than the speed of light in the material, as can be seen in Fig.~(\ref{figures:fig2}). On the other hand, the beat note of the copropagating pump and signals, coupled to the modes with different frequencies, but identical azimuthal quantum numbers, has infinitely large group velocity. These unique properties cannot be easily obtained in an atomic media, but can be observed in a WGM cavity \cite{savchenkov07pra}. Phase matching becomes possible because of the versatility of spectrum engineering in a WGMR.

\vspace{1 mm}

\noindent {\bf Kerr induced laser correlation.}

Now we explore the interplay between Kerr nonlinearity of the resonator and the two counterpropagating lasers injection locked to two of its modes.  We show that phase matching conditions induce formation of phase correlation and frequency locking between the pump lasers and the generated harmonics. This phenomenon is physically similar to the frequency locking observed in slow light resonant FWM atomic systems \cite{fleischhauer00prl}. In our case, when phase matching is achieved the frequency difference between the copropagating pump and signal waves locks to the corresponding frequency of the resonant nonlinear media.  

Consider two counterpropagating intracavity signal ($A_{s1}$ and $A_{s2}$) and pump ($B_{p1}$ and $B_{p2}$) waves resulting from pumping the cavity with two {\em independent } lasers. For the sake of simplicity we assume that the optical modes have zero internal loss and $\gamma_0$ coupling rate. In this fully symmetric case we obtain conditions for the intracavity pump amplitudes above the oscillation threshold
\begin{equation}  \label{saturation} 4g^2|B_{p1}|^2|B_{p2}|^2=\gamma_0^2+\Delta_{p1}\Delta_{p2},
\end{equation}
where $\Delta_{p1}$ and $\Delta_{p2}$ stand for detunings between pump laser frequencies ($\omega_{p1}$ and $\omega_{p2}$) and cross/self phase modulation shifted cavity modes ($\omega_{p10}$ and $\omega_{p20}$)
\begin{eqnarray} \nonumber 
\Delta_{p1}=\omega_{p10}-\omega_{p1}-2g\sum \limits_{i=1}^2 (|A_i|^2+|B_i|^2)+g|A_{p1}|^2, \\ \nonumber 
\Delta_{p2}=\omega_{p20}-\omega_{p2}-2g\sum \limits_{i=1}^2 (|A_i|^2+|B_i|^2)+g|A_{p2}|^2.
\end{eqnarray}
Here $g$ is the nonlinear coefficient of the cavity. Equation (\ref{saturation}) shows that the intracavity pump power saturates after the oscillation threshold is reached. For instance, for the case of zero frequency detunings it can be clearly seen that the product of amplitudes of the pumps becomes a constant in the case of the zero frequency detunings. The detunings can be zeroed down in the case of fast locking of the laser frequency to the corresponding cavity modes. The locking can be facilitated by the self-injection locking process. Therefore, perfect laser locking ensures ideal correlation between the intracavity pump powers. 

When phase matching conditions are satisfied, the frequency difference of the signal produced by the oscillators is locked to frequencies of the cavity modes that are modified by cross-phase modulation. 
\begin{equation} \label{freqdiff}
    \omega_{s1}-\omega_{s2} = \omega_{s10}-\omega_{s20}+g(|A_{s1}|^2-|A_{s2}|^2). 
\end{equation}
In this configuration the laser frequencies can change arbitrarily with respect to the cavity modes, but the frequency difference in (\ref{freqdiff}) holds stable, to the degree of change of the nonlinear frequency shift. Equation (\ref{saturation}) shows that the intracavity powers are correlated due to the oscillations, but the correlation is uncertain because detunings $\Delta_{1,2}$ depend on the laser frequencies. 

The situation changes when the detunings are partially fixed or, in other words, the lasers are locked to the corresponding cavity modes. In a laser-cavity system the laser eigenfrequency does not coincide with the frequency emitted by the laser-cavity system. When the lasers with eigenfrequencies $\omega_{L1}$ and $\omega_{L2}$ are self-injection locked, we obtain the system emission (pump) frequencies
\begin{eqnarray} \label{sil}
    \omega_{p2}-\omega_{p1}= \frac{\omega_{L2}-\omega_{L1}}{1+K}+ \\ \nonumber \frac{K}{1+K} \left [  \omega_{p20}-\omega_{p10} +g (|B_{p2}|^2-|B_{p1}|^2) \right ],
\end{eqnarray}
where $K$ is the self-injection locking coefficient \cite{kondratiev23fp}. For large values of $K$, Eq. (\ref{sil}) describes a pulling of the laser frequencies towards the cross-phase modulation shifted frequency difference of the corresponding nonlinear cavity modes. In this case, the frequency differences between the pumps and signals are locked to the cavity modes, similar to the frequency pulling observed in atomic systems \cite{fleischhauer00prl}.  Here, the frequency pulling phenomenon is enabled by the SIL process; it is not supported in free-running lasers. 

As a result of the frequency pulling effect, the product of the intracavity powers also becomes stabilized, according to Eq~(\ref{saturation}). This means the modulation of power of one of the pump lasers should be efficiently transferred to modulation of power of the other laser. We observed this correlation between the modulation of the pump powers.

\vspace{1 mm}

\noindent {\bf Two-photon oscillation.}

\begin{figure}[htbp]
  \centering
\includegraphics[width=8.5cm]{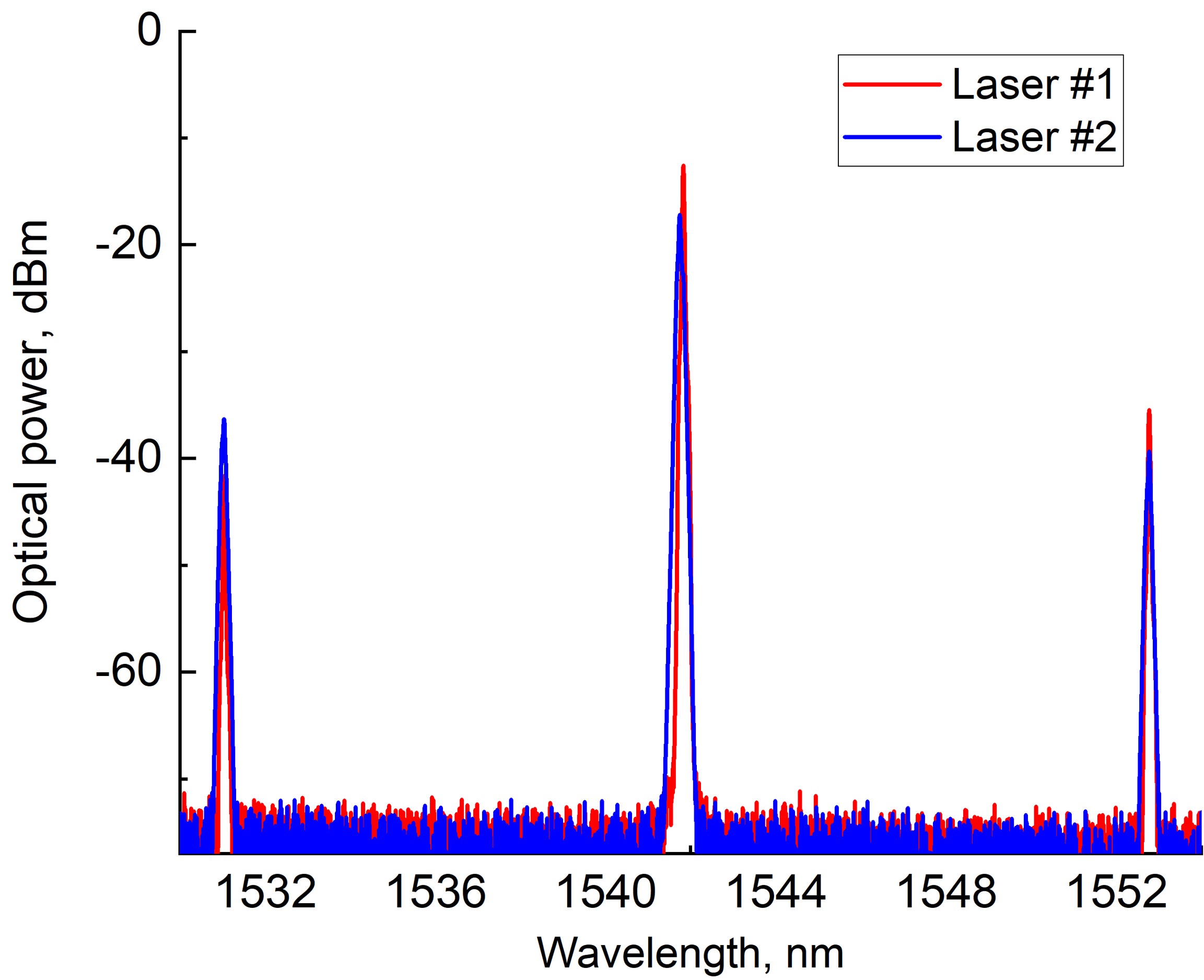}
\captionsetup{singlelinecheck=off, justification = Justified}
\caption{{\bf FWM spectra. } Experimentally observed optical spectra in a nonlinear optical WGM cavity pumped with two counter-propagating coherent optical beams. The carrier frequencies of the pump beams are tuned to different mode families and are separated by 10.7 GHz, which is smaller than the 37 GHz free spectral range of the resonator. While the nonlinear process follows the phase-matching conditions described above, harmonic generation is observed in both directions for each generated harmonic. The unwanted emission is effectively suppressed by approximately an order of magnitude.  \label{figures:fig5}}
\end{figure}

To experimentally demonstrate the oscillation resulting from phase matching of the counter-propagating pumps  we created a setup that incorporated two SIL lasers locked to two modes of a WGM resonator (see Methods). The two lasers could be self-injection locked to two modes of different or the same families by tuning their frequencies  via the current driving the laser chips. In the spectrum shown in Fig.~(\ref{figures:fig5}), the frequency difference of the lasers at 10.7 GHz
is smaller than 30\% of the resonator FSR at 37~GHz. 

Generation of non-degenerate FWM sidebands was achieved by first injection locking one of
the lasers to a target mode and then repeating the process with the second laser. Once injection
locking was accomplished, the power and the phase of light from each laser could be varied, and depending on the values selected, various sidebands were generated and detected with two
optical spectrum analyzers. For  generation of the particular sidebands shown in Fig.~(\ref{figures:fig5}), first the current of laser \#1 was set to 250~mA and then the current in laser \#2 was set to 175 mA. The
temperature of the resonator, controlled by a thermo-electric cooler (TEC), was set at $22^{o}$C. Under these conditions, including a slight adjustment of the phase of one of the lasers, the two sidebands
were generated. To our knowledge, this is the first demonstration of parametric oscillations in a nonlinear WGM (including ring) resonator with two nondegenerate counter-propagating pumps.

Our observations confirm validity of  phase-matching condition for counter-propagating pump waves and demonstrate the feasibility of efficient generation of counter-propagating signal and idler beams within the WGM resonator. Although this process exhibited significant asymmetry, beams were still generated in both directions. According to the phase-matching condition, the sideband emission should ideally exhibit a well-defined directionality. The observed discrepancy can be attributed to practical imperfections in the system, such as Rayleigh scattering, which disrupts the ideal conditions and leads to deviations from the expected direction of emission.

\vspace{1 mm}

\noindent {\bf Laser power correlation induced by the SIL.}

One of the key anticipated characteristics of the observed parametric oscillation is the correlation between the generated sidebands, a phenomenon that has been extensively studied and confirmed in multiple publications. However, a less intuitive, yet significant, aspect is the emergence of correlation between the pump waves themselves. This correlation arises as a consequence of their participation in the joint oscillation process within the system.

In conventional configurations, where the pump sources are independent of the nonlinear medium, such correlation between pumps is not expected. It appears here, however, because each act of the emission of the two photons into the sidebands results in attenuation of a single photon in each of the pumps. Moreover, in our experiments, the lasers are self-injection locked to the resonator cavity. The self-injection locking mechanism creates a strong and fast feedback loop in which the cavity, along with the harmonics generated via the Kerr induced FWM process, influences both pump lasers. As a result, the pumps cannot remain independent but instead become intrinsically correlated through their interaction with the cavity and the nonlinear dynamics of the system.

We checked for evidence for correlation between the two lasers when they were both self-injection locked and under conditions when FWM sidebands were either present or
absent. The current of laser \#1 could be modulated while that of laser \#2 was left untouched. Since both lasers were locked to their respective modes, the current modulation barely could change the frequency of the laser, but produced amplitude modulation.

To observe the effect of modulation of one laser on the output of the other, part of each of the two outputs of the resonator was split and fed to a semiconductor optical amplifier (SOA) and to one of the channels of an oscilloscope. As shown in Figure (\ref{figures:fig3}), the laser
currents could be modulated by the output of a signal generator individually.

\begin{figure}[htbp]
  \centering
\includegraphics[width=8.5cm]{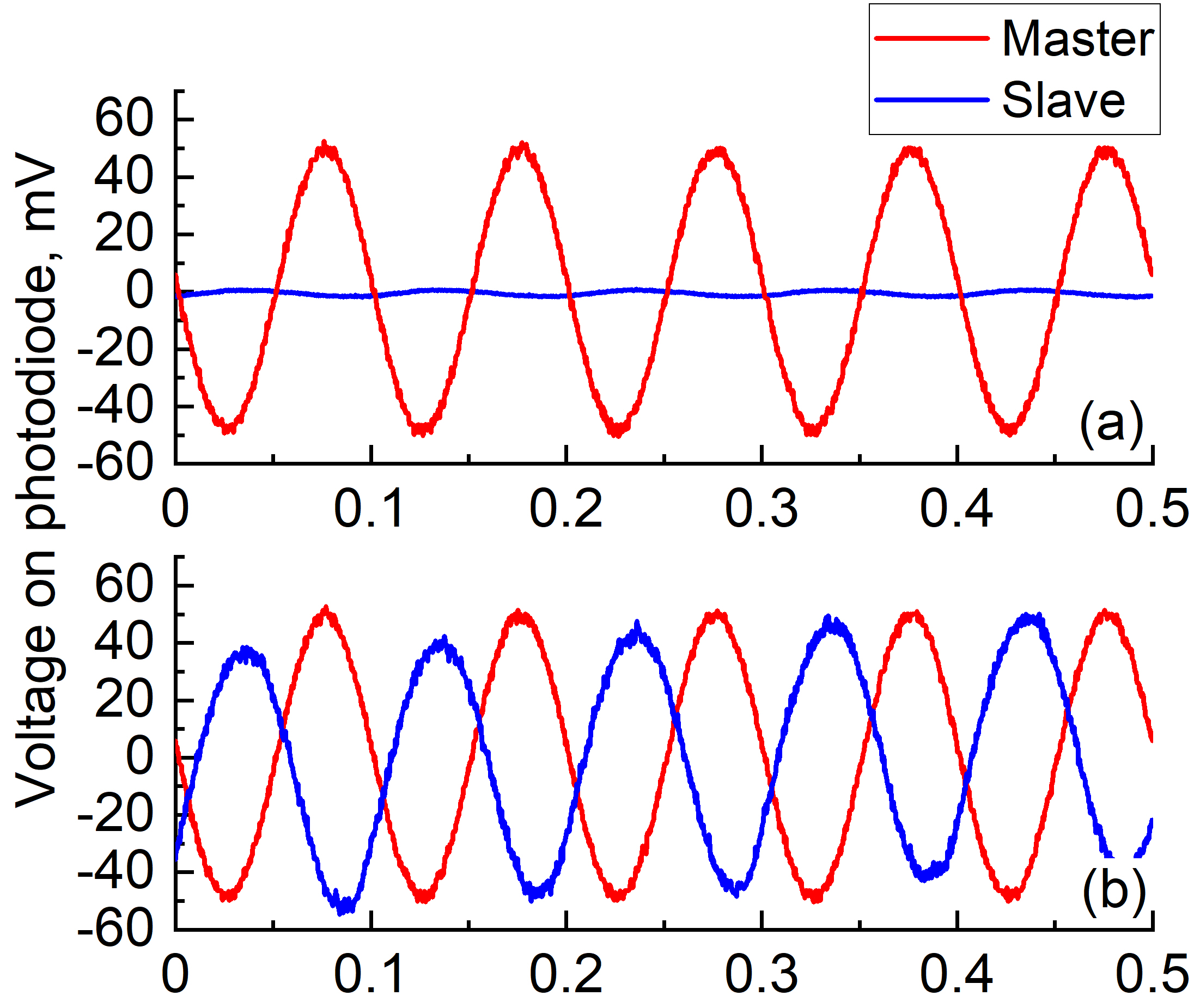}
\captionsetup{singlelinecheck=off, justification = Justified}
\caption{{\bf SIL-induced correlation of pumps.} Transfer of the modulation below and above the parametric oscillation threshold. Power of the laser \#1 (red line) is modulated via its current using a signal generator (Fig.~\ref{figures:fig3}). The frequency modulation is suppressed via the self-injection locking mechanism. The modulation amplitude is measured using a slow photodiode and an oscilloscope. the impact of the amplitude modulation of laser \#1 on the amplitude of the laser \#2 is also detected and is shown by blue line. (a) The lasers operate below the threshold for FWM oscillation. (b) The lasers generate  FWM harmonics, as shown in Fig.~(\ref{figures:fig5}). The four-wave mixing process dramatically produces correlation between thevalues of lasers' power. \label{figures:fig6}}
\end{figure}

\begin{figure}[htbp]
  \centering
\includegraphics[width=8.5cm]{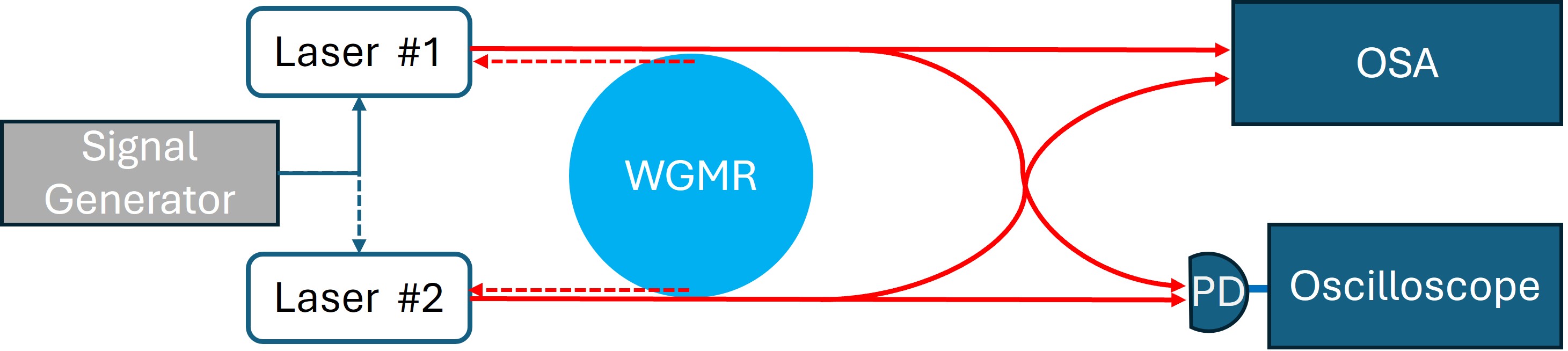}
\captionsetup{singlelinecheck=off, justification = Justified}
\caption{{\bf Experimental setup.} This schematic illustrates the experimental setup. Two semiconductor lasers are each self-injection locked to two counter-propagating modes of the same microcavity, specifically a whispering gallery mode resonator (WGMR). This self-injection locking mechanism arises due to resonant Rayleigh backscattering within the microresonator modes. The eigenfrequency of the lasers can be precisely controlled using two external current sources. The light emerging from the resonator is analyzed using an optical spectrum analyzer (OSA) to examine its spectral properties, while a photodiode (PD) and an oscilloscope are used to investigate its temporal characteristics and signal dynamics. \label{figures:fig3}}
\end{figure}

In this way, we could observe both the appearance or absence of degenerate FWM sidebands, and the effect of current modulation on each laser. When the selected values of optical power for each SIL laser were set below the threshold of FWM oscillation where no sidebands were produced, modulation of laser \#1 had almost no effect on the output of laser \#2, as seen in Figure (\ref{figures:fig6}a). But when the system was set at parameters that generated FWM sidebands, modulation of the current of laser \#1 produced a strong modulated output on laser \#2 (Fig.\ref{figures:fig6}b). This effect was repeatable when the roles of laser \#1 and laser \#2 as current modulation sources were exchanged. To our knowledge, this is the first observation of  correlation between two pump lasers producing FWM and represents yet another important result of our study with the configuration of two SIL lasers counter-propagating in a WGMR. 

\medskip

\noindent {\bf Conclusion. } 
In this work we have studied a new configuration for investigation of interaction of two lasers and their counterpropagating optical fields with mediated by the Kerr nonlinearity of a WGMR.  The configuration consists of the lasers injection locked  to two different modes of the resonator.  Non-degenerate four-wave mixing  was both theoretically predicted and experimentally demonstrated in the high quality factor nonlinear WGMR. This, to the best of our knowledge, is the first demonstration of non-degenerate four-wave-mixing in a WGM (including ring) resonator.  

Another significant outcome of this process was the first experimentally observed FWM-induced correlation between the two optical pumping lasers, a phenomenon that highlights the intricate nonlinear interaction occurring within the resonator in this configuration.

We show that a crucial factor enabling this demonstration is the role of self-injection locking of the independent laser chips to the modes of the same optical microcavity. These laser chips are hybridly integrated onto a shared micro-platform, making them highly suitable for scalable fabrication and practical implementation. The self-injection locking mechanism not only stabilizes the lasers but also facilitates the strong nonlinear interaction required for efficient FWM threshold.

Furthermore, as shown in several previous studies, the counter-propagating optical fields generated in this system are expected to exhibit quantum correlations. This suggests the potential of this approach for applications in quantum optics and precision metrology, where nonclassical correlations  are harnessed for enhanced performance in tasks such as quantum-enhanced sensing and secure communication.

Finally, while the work reported here was performed with an assembly consisting of free-space micro-optical elements, the architecture is readily suitable for on-chip integration on a photonic integrated circuit, PIC.

\vspace{1 mm}
\noindent \textbf{Acknowledgments}
The research reported here performed by A.M. was carried out at the Jet Propulsion Laboratory at the California Institute of Technology, under a contract with the National Aeronautics and Space Administration (80NM0018D0004).  The research performed at OEwaves was partially funded by Air Force STTR 2022—SF22D-T005and, and in part by OEwaves internal IR\&D.

\vspace{1 mm}

\noindent\textbf{Author Contributions:} Concepts were developed by L.M. and A.M.  LM designed the experimental setup and A.E.A helped with its assembly. A.E.A. and L.M. performed the experiments.  AM developed the theory and L.M. and A.M. analyzed the data. AM and LM contributed to the writing of the manuscript. 
\vspace{1 mm}

\noindent \textbf{Competing Interests:} Lute Maleki and Abdelkarim El Amili are with OEwaves Inc. Andrey Matsko declares no competing interests. Abdelkarim El Amili is currently with Advanced Products Research Lab.
\vspace{1 mm}

\noindent \textbf{ Data availability:} Data sets generated during the current study are available from the corresponding author on reasonable request.

\clearpage

\vspace{3 mm}
\noindent\textbf{Methods}

\begin{figure}[htbp]
  \centering
\includegraphics[width=8.5cm]{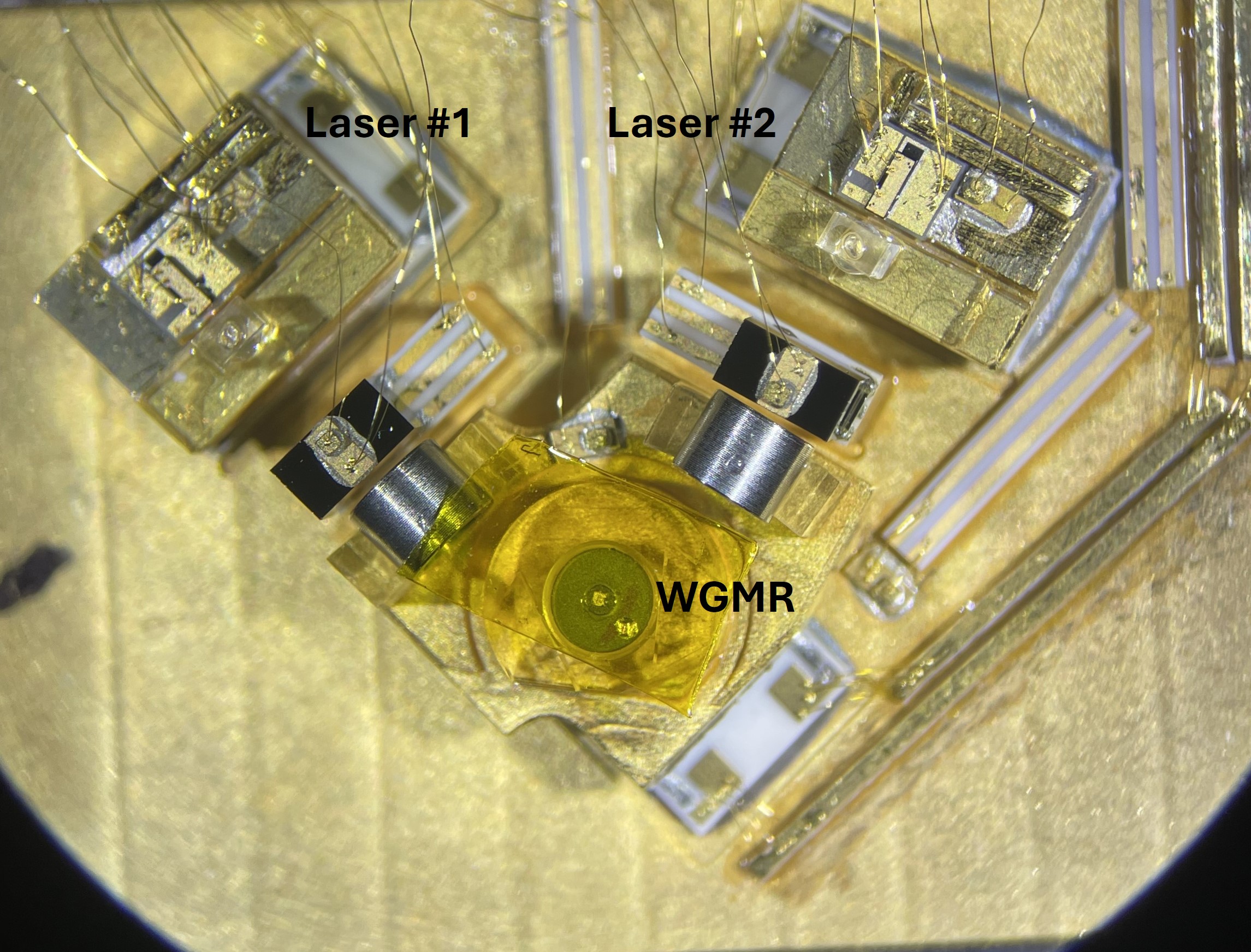}
\captionsetup{singlelinecheck=off, justification = Justified}
\caption{{\bf Physics package.} This image depicts the nonlinear microresonator section of the experimental setup.  Two semiconductor lasers are configured to generate counter-propagating light into the WGMR.  Light from each laser is collimated and directed each into a different whispering gallery mode of the resonator  using two evanescent field prism couplers. Through the resonant Rayleigh backscattering effect, the lasers become self-injection locked to the corresponding resonator mode, which enhance their coherence and stability. The light emerging from the resonator in two directions is collected using two standalone optical fibers, which are not visible in the image.  \label{figures:fig4}}
\end{figure}

\noindent {\bf Experimental setup.} A bright source of photon pairs was generated via FWM through an experimental setup consisting of free-space micro-optical components.  It featured two semiconductor chip lasers with emission wavelength of approximately 1549.7 nm self-injection locked to a WGMR (See Figures~\ref{figures:fig3} and \ref{figures:fig4}). The magnesium fluoride resonator, with a diameter of 2.8 mm (free spectral range, FSR, of 37~GHz), was fabricated using grinding and polishing techniques. The fundamental mode of the resonator exhibited a quality factor (Q) of $1.2\times  10^9$.  

Each laser beam was first collimated using a lens before being directed toward the input face of a coupling prism. These coupling prisms were strategically positioned near the edge of the WGMR to ensure efficient evanescent coupling while preserving the high-Q of the system. The placement was carefully optimized so that the evanescent fields of the prisms overlapped with the evanescent field of the circulating light within the resonator, enabling efficient energy transfer without introducing excessive losses.

The experimental setup was designed to facilitate counter-propagating optical pumping, meaning that the laser inputs were arranged to propagate in opposite directions within the resonator. This configuration is essential for investigating nonlinear interactions such as FWM and for exploring the resulting correlations between the counter-propagating optical fields. By ensuring precise alignment and maintaining the resonator’s high Q-factor, the system was optimized for robust and efficient nonlinear optical interaction.

The two counter-propagating optical outputs from the resonator were collected using separate light collection systems, each positioned near the output face of its corresponding coupling prism. Each collection system consisted of a fiber optic cable that could be connected either directly to an optical spectrum analyzer (OSA) for spectral analysis or to a photodetector for real-time signal monitoring.

Additionally, the fiber optic outputs from both collection systems could be combined using a fiber coupler, allowing the signals to be routed to a single photodetector for joint analysis. The electrical signals from each photodetector were then fed into an oscilloscope, enabling real-time observation of the resonator’s output dynamics. This configuration provided flexibility, allowing the outputs to be simultaneously monitored on two independent optical spectrum analyzers or a multi-channel oscilloscope, facilitating a comprehensive analysis of the system’s  behavior.

To maintain stable operation and achieve self-injection locking of the lasers to the optical microcavity, the temperature and current of each laser were independently controlled. Precise regulation of these parameters ensured that the lasers remained locked to the resonator modes, optimizing the efficiency of nonlinear interactions such as FWM and enabling robust experimental investigations.

\vspace{1 mm}

\begin{thebibliography}{99}

\bibitem{matsko06jstqe} Matsko, A.B. and Ilchenko, V.S.,  "Optical resonators with whispering gallery modes I: basics," IEEE J. Sel. Top. Quantum Electron, 12(3), 3-14 (2006).

\bibitem{chiasera10lpr} Chiasera, A., Dumeige, Y., Feron, P., Ferrari, M., Jestin, Y., Nunzi Conti, G., Pelli, S., Soria, S. and Righini, G.C., "Spherical whispering‐gallery‐mode microresonators," Laser \& Photonics Reviews, 4(3), pp.457-482 (2010).

\bibitem{gorodetsky96ol} Gorodetsky, M.L., Savchenkov, A.A. and Ilchenko, V.S., "Ultimate Q of optical microsphere resonators," Optics Letters, 21(7), 453-455 (1996).

\bibitem{savchenkov07oe} Savchenkov, A.A., Matsko, A.B., Ilchenko, V.S. and Maleki, L., "Optical resonators with ten million finesse," Optics Express, 15(11), pp.6768-6773 (2007).

\bibitem{delhaye07n} Del’Haye, P., Schliesser, A., Arcizet, O., Wilken, T., Holzwarth, R. and Kippenberg, T.J.,  "Optical frequency comb generation from a monolithic microresonator," Nature, 450(7173), 1214-1217 (2007).

\bibitem{savchenkov08prl} Savchenkov, A.A., Matsko, A.B., Ilchenko, V.S., Solomatine, I., Seidel, D. and Maleki, L., "Tunable optical frequency comb with a crystalline whispering gallery mode resonator," Phys. Rev. Lett. 101(9), art. no. 093902 (2008).

\bibitem{kippenberg11s} Kippenberg, T.J., Holzwarth, R. and Diddams, S.A., 2011. Microresonator-based optical frequency combs. Science, 332(6029), pp.555-559 (2011).

\bibitem{liang10ol}  Liang, W., Ilchenko, V. S., Savchenkov, A. A., Matsko, A. B., 
Seidel, D., and Maleki, L., "Whispering-gallery-moderesonator-
based ultranarrow linewidth external-cavity semiconductor
laser," Opt. Lett. 35, 2822 (2010).

\bibitem{ohta68ecj} Ohta, T. and Murakami, K., "Reducing negative resistance oscillator noise by self injection," Electron. Commun. Jpn. 51-B, 80 (1968).

\bibitem{dahmani87ol} Dahmani, B.,  Hollberg, L.,  and Drullinger, R., "Frequency
stabilization of semiconductor lasers by resonant optical
feedback," Opt. Lett. 12, 876 (1987).

\bibitem{gorodetsky00josab} Gorodetsky, M. L.,  Pryamikov, A. D., and Ilchenko, V. S.,
"Rayleigh scattering in high-Q microspheres," J. Opt. Soc.
Am. B 17, 1051 (2000).

\bibitem{liang15nc} Liang, W., Ilchenko, V., Eliyahu, D., Savchenkov, A. A.,  Matsko, A. B., Seidel, D., and Maleki, L., "Ultralow noise miniature external
cavity semiconductor laser," Nat. Commun. 6, 7371 (2015).

\bibitem{voloshin21nc} Voloshin, A. S., Kondratiev, N. M., Lihachev, G. V., Liu, J., Lobanov, V. E., Dmitriev, N. Y., Weng, W., Kippenberg, T.J., and Bilenko, I. A.,  "Dynamics of soliton self-injection locking in optical microresonators," Nature Communications, 12(1), art. no. 235 (2021).

\bibitem{maleki11np} Maleki, L., "The optoelectronic oscillator," Nature Photonics, 5(12), 728-730 (2011).

\bibitem{kondratiev23fp} Kondratiev, N.M., Lobanov, V.E., Shitikov, A.E., Galiev, R.R., Chermoshentsev, D.A., Dmitriev, N.Y., Danilin, A.N., Lonshakov, E.A., Min’kov, K.N., Sokol, D.M. and Cordette, S.J., Luo, Y.-H., Liang, W., Liu, J., and Bilenko, I. A., "Recent advances in laser self-injection locking to high-Q microresonators," Frontiers of Physics, 18(2), 21305 (2023).

\bibitem{liang15nc1} Liang, W., Eliyahu, D., Ilchenko, V.S., Savchenkov, A.A., Matsko, A.B., Seidel, D. and Maleki, L., "High spectral purity Kerr frequency comb radio frequency photonic oscillator," Nature Communications, 6(1), art. no. 7957 (2015).

\bibitem{shen20n} Shen, B., Chang, L., Liu, J., Wang, H., Yang, Q.F., Xiang, C., Wang, R.N., He, J., Liu, T., Xie, W. Guo, J.,  Kinghorn, D., Wu, L., Ji, Q.-X., Kippenberg, T. J., Vahala, K., Bowers, J. E., "Integrated turnkey soliton microcombs," Nature, 582(7812), pp.365-369 (2020).

\bibitem{geng22ol} Geng, Y., Xiao, Y., Bai, Q., Han, X., Dong, W., Wang, W., Xue, J., Yao, B., Deng, G., Zhou, Q. and Qiu, K., "Wavelength-division multiplexing communications using integrated soliton microcomb laser source," Opt. Lett. 47(23), pp.6129-6132 (2022).

\bibitem{pisque19np} Picque, N. and Hänsch, T.W.,"Frequency comb spectroscopy," Nature Photonics, 13(3), pp.146-157 (2019).

\bibitem{savchenkov20sr} Savchenkov, A.A., Christensen, J.E., Hucul, D., Campbell, W.C., Hudson, E.R., Williams, S. and Matsko, A.B., "Application of a self-injection locked cyan laser for barium ion cooling and spectroscopy," Scientific Reports, 10(1), p.16494 (2020).

\bibitem{voloshin24arxiv} Voloshin, A., Siddharth, A., Bianconi, S., Attanasio, A., Bancora, A., Shadymov, V., Leni, S., Wang, R.N., Riemensberger, J., Bhave, S.A. and Kippenberg, T.J., "Monolithic piezoelectrically tunable hybrid integrated laser with sub-fiber laser coherence," arXiv preprint arXiv:2411.19264 (2024).

\bibitem{heim25arxiv} Heim, D.A., Bose, D., Liu, K., Isichenko, A. and Blumenthal, D.J., "Hybrid integrated ultra-low linewidth coil stabilized isolator-free widely tunable external cavity laser," arXiv preprint arXiv:2501.15010 (2025).

\bibitem{chang22np} Chang, L., Liu, S. and Bowers, J.E., Integrated optical frequency comb technologies. Nature Photonics, 16(2), 95-108 (2022).

\bibitem{hansson14pra} Hansson, T. and Wabnitz, S., “Bichromatically pumped microresonator frequency combs,” Phys. Rev. A, 90(1),  013811 (2014).

\bibitem{wang16sr} Wang, W., Chu, S.T., Little, B.E., Pasquazi, A., Wang, Y., Wang, L., Zhang, W., Wang, L., Hu, X., Wang, G. and Hu, H., “Dual-pump Kerr micro-cavity optical frequency comb with varying FSR spacing. Scientific Reports, 6(1),  28501 (2016).

\bibitem{suzuki18pj} Suzuki, R., Fujii, S., Hori, A. and Tanabe, T., “Theoretical study on dual-comb generation and soliton trapping in a single microresonator with orthogonally polarized dual pumping,” IEEE Photonics J. 11(1), pp.1-11 (2018).

\bibitem{taheri22nc} Taheri, H., Matsko, A.B., Maleki, L. and Sacha, K., “All-optical dissipative discrete time crystals,” Nature Communications, 13(1),  848 (2022).

\bibitem{wildi23apl}  Wildi, T., Ulanov, A., Englebert, N., Voumard, T. and Herr, T., “Sideband injection locking in microresonator frequency combs,” APL photonics, 8(12), 120801 (2023).

\bibitem{moille24np} Moille, G., Leonhardt, M., Paligora, D., Englebert, N., Leo, F., Fatome, J., Srinivasan, K. and Erkintalo, M., “Parametrically driven pure-Kerr temporal solitons in a chip-integrated microcavity,” Nature Photonics 18(6), pp.617-624 (2024).

\bibitem{jiang21oe} Jiang, L., Shi, L., Luo, J., Gao, Q., Bai, M., Lan, T., Iroegbu, P. I., Dang, L., Huang, L., Zhu, T., “Simultaneous self-injection locking of two VCSELs to a single whispering-gallery-mode microcavity,” Opt. Express 29(23), 37845–37851 (2021).

\bibitem{chermoshentsev22oe} Chermoshentsev, D.A., Shitikov, A.E., Lonshakov, E.A., Grechko, G.V., Sazhina, E.A., Kondratiev, N.M., Masalov, A.V., Bilenko, I.A., Lvovsky, A.I. and Ulanov, A.E., “Dual-laser self-injection locking to an integrated microresonator.” Optics Express, 30(10), pp.17094-17105 (2022). 

\bibitem{chembo16pra} Chembo, Y.K., "Quantum dynamics of Kerr optical frequency combs below and above threshold: Spontaneous four-wave mixing, entanglement, and squeezed states of light," Physical Review A, 93(3), 033820 (2016).

\bibitem{zhao20prl} Zhao, Y., Okawachi, Y., Jang, J.K., Ji, X., Lipson, M. and Gaeta, A.L., “Near-degenerate quadrature-squeezed vacuum generation on a silicon-nitride chip,” Physical Review Letters, 124(19), 193601 (2020).

\bibitem{okawachi15ol} Okawachi, Y., Yu, M., Luke, K., Carvalho, D.O., Ramelow, S., Farsi, A., Lipson, M. and Gaeta, A.L., 2015. Dual-pumped degenerate Kerr oscillator in a silicon nitride microresonator. Optics letters, 40(22), pp.5267-5270 (2015). 

\bibitem{hu17pr} Hu, X., Wang, W., Wang, L., Zhang, W., Wang, Y. and Zhao, W., “Numerical simulation and temporal characterization of dual-pumped microring-resonator-based optical frequency combs,” Photonics Research, 5(3), pp.207-211, (2017).

\bibitem{seifoory22pra} Seifoory, H., Vernon, Z., Mahler, D.H., Menotti, M., Zhang, Y. and Sipe, J.E., “Degenerate squeezing in a dual-pumped integrated microresonator: Parasitic processes and their suppression,” Physical Review A, 105(3), p.033524 (2022).

\bibitem{marty21np} Marty, G., Combrie, S. , Raineri, F., and De Rossi, A., “Photonic crystal optical parametric oscillator,” Nature Photon., vol. 15, no. 1, pp. 53–58 (2021).

\bibitem{dutt15pra} Dutt, A., Luke, K., Manipatruni, S., Gaeta, A.L., Nussenzveig, P. and Lipson, M., 2015. On-chip optical squeezing. Physical Review Applied, 3(4), 044005 (2015).

\bibitem{matsko16ol} Matsko, A.B., Savchenkov, A.A., Huang, S.W. and Maleki, L., 2016. Clustered frequency comb. Optics Letters, 41(21), pp.5102-5105 (2016).

\bibitem{shou19lsa} Zhou, H., Geng, Y., Cui, W., Huang, S.W., Zhou, Q., Qiu, K. and Wei Wong, C., “Soliton bursts and deterministic dissipative Kerr soliton generation in auxiliary-assisted microcavities,” Light: Science \& Applications, 8(1), p.50 (2019).

\bibitem{delbino17sr} Del Bino, L., Silver, J.M., Stebbings, S.L. and Del'Haye, P., “Symmetry breaking of counter-propagating light in a nonlinear resonator,” Scientific Reports, 7(1), p.43142 (2017).

\bibitem{yoshiki15oe} Yoshiki, W., Chen-Jinnai, A., Tetsumoto, T. and Tanabe, T., “Observation of energy oscillation between strongly-coupled counter-propagating ultra-high Q whispering gallery modes,” Optics Express, 23(24), pp.30851-30860 (2015).

\bibitem{matsko17ol} Matsko, A. B., and Maleki, L., "Bose–Hubbard hopping due to resonant Rayleigh scattering," Opt. Lett. 42, 4764-4767 (2017).

\bibitem{RodríguezBecerra24ao} Rodríguez Becerra, G.J., Durán Gómez, J.S.S., Tavares Ramírez, P.M.C., Ramírez Alarcón, R., Gómez Robles, M. and Salas-Montiel, R., “Integrated photon pairs source based on counter-propagating spontaneous four wave mixing in a silicon nitride microring resonator,”Applied Optics, 63(27), pp.7278-7285 (2024).

\bibitem{zibrov99prl} Zibrov, A.S., Lukin, M.D. and Scully, M.O., "Nondegenerate parametric self-oscillation via multiwave mixing in coherent atomic media," Physical Review Letters, 83(20), 4049 (1999).

\bibitem{liu19prl} Liu, S., Lou, Y. and Jing, J.,  "Interference-induced quantum squeezing enhancement in a two-beam phase-sensitive amplifier," Physical Review Letters, 123(11), 113602 (2019).

\bibitem{mccormick08pra} McCormick, C.F., Marino, A.M., Boyer, V. and Lett, P.D., "Strong low-frequency quantum correlations from a four-wave-mixing amplifier," Physical Review A, 78(4), 043816 (2008).

\bibitem{pooser09oe} Pooser, R. C., Marino, A. M., Boyer, V., Jones, K. M., Lett, P. D., “Quantum correlated light beams from  nondegenerate four-wave mixing in an atomic vapor: the D1 and D2 lines of 85Rb and 87Rb,” Opt. Express, Vol. 17, No.  19, 16722 (2009). 

\bibitem{sumetsky04ol}  Sumetsky, M., “Whispering-gallery bottle microcavities: the three-dimensional etalon,” Opt. Lett. 29, 8–10 (2004).

\bibitem{demchenko13josab}  Demchenko, Y. A., Gorodetsky, M. L., "Analytical estimates of eigenfrequencies, dispersion, and field distribution in whispering gallery resonators," J. Opt. Soc. Am. B 30, 3056-3063 (2013).

\bibitem{savchenkov07pra} Savchenkov, A.A., Matsko, A.B., Ilchenko, V.S., Strekalov, D. and Maleki, L., "Direct observation of stopped light in a whispering-gallery-mode microresonator," Physical Review A 76(2), 023816 (2007).

\bibitem{fleischhauer00prl} Fleischhauer, M., Lukin, M.D., Matsko, A.B. and Scully, M.O., "Threshold and linewidth of a mirrorless parametric oscillator," Physical Review Letters, 84(16), 3558 (2000).

\end{thebibliography}
\end{document}